# Wall Stress Distributions in Abdominal Aortic Aneurysms
# Do Not Correlate With Symptoms

K. Miller[1], H. Mufty[2], A. Catlin[1], C. Rogers[1], B. Saunders[1], R. Sciarrone[1], I. Fourneau[2],
B. Meuris[3], A. Tavner[1], G. R. Joldes[1], A. Wittek[1*]

[1]Intelligent Systems for Medicine Laboratory, The University of Western Australia,
Perth, Western Australia

[2]Vascular Surgery, University Hospitals Leuven, Belgium

[3]Cardiac Surgery, University Hospitals Leuven, Belgium

*Corresponding author: E-mail: adam.wittek@uwa.edu.au

**Abstract:**

Abdominal aortic aneurysm (AAA) is a permanent and irreversible dilation of the lower region of the aorta. It is typically an asymptomatic condition that if left untreated can expand to the point of rupture. Mechanically-speaking, rupture of an artery occurs when the local wall stress exceeds the local wall strength. It is therefore understandable that numerous studies have attempted to estimate the AAA wall stress. Recently the Intelligent Systems for Medicine Laboratory (ISML) presented a very efficient method to compute AAA wall stress using geometry from Computed Tomography (CT) images, and median arterial pressure as the applied load. The ISML's method is embedded in the software platform *BioPARR - Biomechanics based Prediction of Aneurysm Rupture Risk*, freely available from http://bioparr.mech.uwa.edu.au/. The uniqueness of our stress computation approach is three-fold: i) the results are insensitive to unknown patient-specific mechanical properties of arterial wall tissue; ii) the residual stress is accounted for, according to Y.C. Fung's Uniform Stress Hypothesis; and iii) the analysis is automated and quick, making our approach compatible with clinical workflows. In this study we evaluated 19 cases of AAA. A proportion of these were classified as symptomatic. The results of the analysis demonstrate, contrary to the common view, that neither the wall stress magnitude nor the stress distribution correlate with the presence of clinical symptoms.

# 1    Introduction

Abdominal aortic aneurysm (AAA) is a permanent and irreversible dilation of the lower region of the aorta, is typically asymptomatic, and if untreated can result in rupture of the aorta. AAA is found in approximately 7% of elderly men (>65 yrs) in Australia (Norman et al., 2004) with similar prevalence throughout the Western world (Singh et al., 2001). The disease also affects women, but at a lower rate.

Because AAA is usually asymptomatic, most people are unaware of their condition. However, AAA rupture is a catastrophic clinical event with mortality rates of more than 80% (Badger et al., 2016; Bengtsson and Bergqvist, 1993; Evans et al., 2000; Kantonen et al., 1999). Currently, the most widely-used evidence-based indicator of rupture threat (based on several large clinical trials) is the maximum anterior-posterior diameter: diameters greater than 5.5cm in men, diameters greater than 5.0cm in women or expansion rates of greater than 1.0cm over the preceding 12 months are deemed high risk (Wanhainen et al 2019). These cut-off values must be weighed against associated co-morbidities and intra-operative mortality risk, with repair only being considered if the risk of rupture exceeds the risk of surgery. However 20% of smaller AAAs rupture, while larger cases often remain quiescent (Darling et al., 1977; Greenhalgh et al., 2004). The ability to predict, non-invasively, which cases are at risk of rupture will have a major clinical impact by saving lives and reducing medical costs worldwide.

When patients present with symptoms, more urgent action is required; in the case of a ruptured AAA (rAAA) showing clearly visible signs on computed tomography angiography (CTa) immediate intervention will be required. Patients presenting with symptoms but no rupture apparent on CTa pose more of a challenge. These patients present with abdominal pain, back pain or symptoms from local compression caused by the aneurysm (i.e. hydronephrosis, deep vein thrombosis and early satiety) and a clinical judgment must be made about whether emergency surgery is required.

Because of the limitations of the current clinical definition of 'high-risk', many researchers believe that patient-specific modelling (PSM) has significant clinical potential (Gasser et al., 2010; Gasser et al., 2014; Joldes et al., 2016; McGloughlin and Doyle, 2010; Vande Geest et al., 2006; Zelaya et al., 2014). In simple mechanical terms, rupture of an artery will occur when the local wall stress exceeds the local wall strength. With advances in medical imaging technology and medical image analysis software, it has become possible to create patient-specific reconstructions of the AAA, which can then be used for computer simulations aimed at computing the wall stress.   These models have steadily increased in complexity (Doyle et al., 2007; Gasser et al., 2010; Li et al., 2010; Raghavan et al., 2000). Major research efforts have been preoccupied with material models and simulations so comprehensive that they require computing resources and specialist expertise that are not likely to be available in a typical clinical setting.

A simple approach to compute AAA wall stress was recently proposed and validated by Joldes et al. (2016) in *Journal of the Mechanical Behavior of Biomedical Materials* (but see also Biehler et al., 2015; Fung, 1991; Zelaya et al., 2014). The inputs to the model are the (loaded) geometry of an aneurysm (obtained from a CT reconstruction), wall thickness and blood pressure. Our approach also efficiently incorporates residual stresses according to Fung's Uniform Stress Hypothesis (Fung, 1991; Joldes et al., 2018). The method is embedded in the software platform *BioPARR -*

*Biomechanics based Prediction of Aneurysm Rupture Risk* (Joldes et al., 2017), freely available from http://bioparr.mech.uwa.edu.au/. This approach does not require any information about arterial tissue material parameters, which supports the development and use of PSM, where uncertainty in material data, until relatively recently (Miller and Lu, 2013), has been regarded as a key limitation. Furthermore, the computation itself is so simple that incorporating it into existing clinical workflows does not represent a significant challenge. In this study, we aim to investigate the use of BioPARR in predicting AAA rupture risk in symptomatic and asymptomatic patients. There are suggestions in the literature, for example (Fillinger et al., 2003; Vorp, 2007) and a comprehensive review (Khosla et al., 2014), that patients presenting with symptoms will have high stresses in the arterial wall of an AAA. Therefore, we have applied our methodology to investigate whether the wall stress fields (both the patterns and magnitude) correlate with the clinical definition of symptomatic and asymptomatic AAAs.

## 1.1 Population

Anonymised data from 25 patients with radiographically clear un-ruptured AAAs were used for this study. All patients were treated in the University Hospitals Leuven, Belgium. Five patients presented with symptoms, 20 patients were asymptomatic. In both groups, patients were selected independent of age, gender or comorbidities. Data from patients were selected retrospectively and were provided to the University of Western Australia Intelligent Systems for Medicine Laboratory. The study was approved by the local ethics committee of the University Hospitals Leuven (approval no. S59796). Patient demographics are listed in Table 1 (overleaf).

**Table 1**: Patient demographics

| Patient demographic | | Value | % |
|---|---|---|---|
| Gender (Male/Female) | | 24/1 | 96% / 4% |
| Age (Mean + range) | | 72y (58-83) | |
| BMI (kg/cm²) (Mean + range) | | 27 (16-33) | |
| Maximum diameter AAA (Mean + range) | | 61mm (37-86) | |
| Pulse (Mean + range) | | 74 bps (51-116) | |
| Blood pressure (mmHg) | | | |
| Mean diastolic BP (range) | | 79 (44-120) | |
| Mean systolic BP (range) | | 145 (110-204) | |
| Mean arterial pressure (range) | | 100 (69-148) | |
| Smoking: (N) | Non-smoker | 5 | 20% |
| | Active smoker | 8 | 32% |
| | Stop > 10 y | 6 | 24% |
| | Stop < 10 y | 6 | 24% |
| Diabetes mellitus (N) | | | |
| | None | 21 | 84% |
| | Type 1 | 0 | 0% |
| | Type 2 | 4 | 16% |
| PAD (N) | Yes | 2 | 8% |
| | No | 23 | 92% |
| Arterial hypertension (N) | | 15 | 60% |
| Hypercholesterolemia (N) | | 20 | 80% |
| Open/endovascular approach (N) | | 10/15 | 40% / 60% |

bps: beats per second; PAD: peripheral arterial disease; N: number of cases

## 2   Material and Methods

Complete stress analyses of each AAA were conducted using our freely-available software BioPARR (Joldes et al., 2017). The 3D reconstruction time of each case including segmentation took approximately 40 minutes, and the numerical analysis of a single load case scenario, including the incorporation of residual stress, took approximately 6 minutes on an Intel(R) Core(TM) i7-5930K CPU @ 3.50GHz with 64GB of RAM running Windows 8 OS. The analysis steps are briefly described below.

### 2.1   Problem geometry

We used 25 real-world, patient-specific 3D geometries of patients, with all the irregularities that can be expected in clinical simulations, Figure 1. However, due to limited image quality, we were not able to conduct reliable segmentation for 6 patients. Therefore, the analysis of stress within AAA wall was done for only 19 patients.

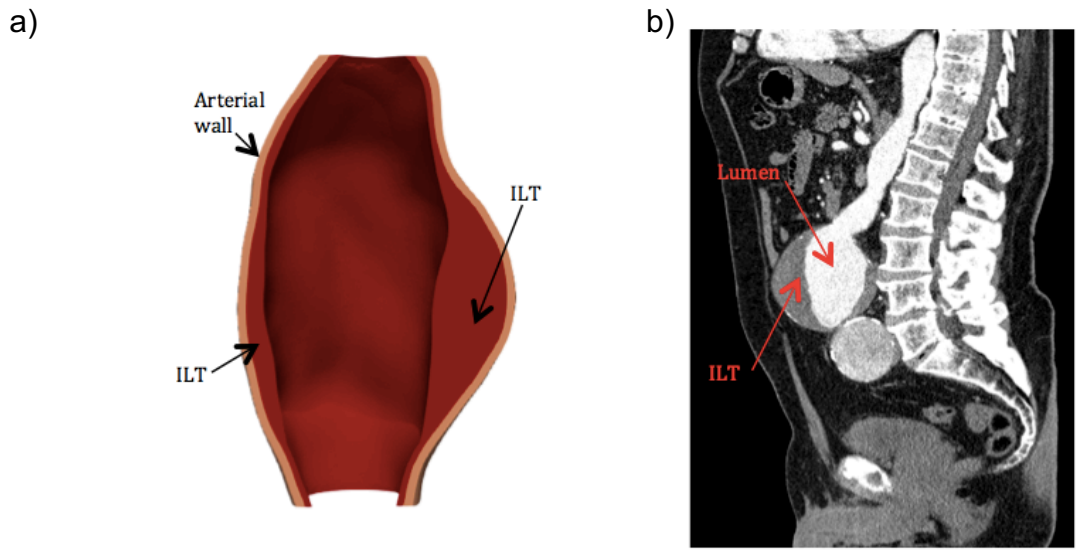

**Figure 1:** An example of AAA case analysed in this study. a) Region of interest, showing the lumen and portions of the AAA wall and intraluminal thrombus (ILT); b) CT image.

BioPARR allows the analyst to extract and combine data from images of different modality (such as CT and MRI), by implementing a segmentation-based inter-modality image registration algorithm in 3D Slicer (Fedorov et al., 2012). The analyst has control over many parameters influencing the analysis results: the thickness of the AAA wall, inclusion of the thrombus, geometry meshing, finite element type selection, and finite element simulation scenarios. The software can be used in the situation when both CT and MRI data are available for a patient or, the more typical situation, when only CT is available. CT images are acquired as a part of routine care and are available for most clinically relevant AAAs.

The program automatically generates 3D colour-contoured visualisations of the key patient-specific components of the analysis, namely, intraluminal thrombus (ILT) thickness and the normalised ratio of the maximum AAA diameter and the diameter in the proximal neck of the aneurysm (NORD).

## 2.2 Image segmentation

The high variability in AAA geometry, as well as low discrimination between the AAA and the surrounding tissue in parts of the image, make automatic AAA segmentation practically impossible. Therefore, our software uses segmentation tools from the freely-available open-source image analysis software 3D Slicer (Fedorov et al., 2012). We have found that using the 3D Slicer extension *FastGrowCut* for segmentation (Zhu et al., 2014) can help reduce the segmentation time. Manual input from the analyst is still required to define the region of interest in the image, to crop the image, and to define the seeds for the *FastGrowCut* algorithm. Manual corrections and smoothing of the resulting label maps are also necessary. Using this method, we can extract the AAA geometry from CT (or MRI if available).

## 2.3 Geometry creation

The label maps segmented from images combined with the assumed wall thickness of 1.5mm (measurement of wall thickness from CT images is not possible) were used to create the AAA geometry for the models. The external AAA wall surface, the internal AAA wall surface and the internal intraluminal thrombus (ILT) surface were automatically created. The AAA wall thickness of 1.5mm we use here has been previously applied in the literature for AAA biomechanics analysis (Raut et al. 2013). However, other values (2mm) have also been reported and used (Khosla et al, 2014). Discussion of the assumption of constant wall thickness in the context of model limitation is provided in *Discussion and Conclusions* section.

## 2.4 Finite element meshing, Model creation and Analysis

Meshing of the AAA wall and ILT, based on the external and internal AAA wall surfaces and the internal ILT surface, was performed using open-source meshing software Gmsh (Geuzaine and Remacle, 2009; Geuzaine and Remacle, 2016) called from within BioPARR. A tetrahedral volumetric mesh was created using the element size specified by the user. This process ensures a conforming mesh between the ILT and AAA wall. The meshing approach implemented in BioPARR uses very small elements on the surfaces to maintain the geometric accuracy. At the same time, by increasing the element size inside the ILT volume and in the AAA wall, the overall mesh size is reduced along with the consequent computational cost of the finite element analysis. The element types can be configured as linear or quadratic, displacement only or hybrid displacement-pressure formulation. Finally, Abaqus (Abaqus, 2018) text input (.inp) files are generated and sent for finite element analysis. Figure 2 shows a typical AAA mesh used in this study.

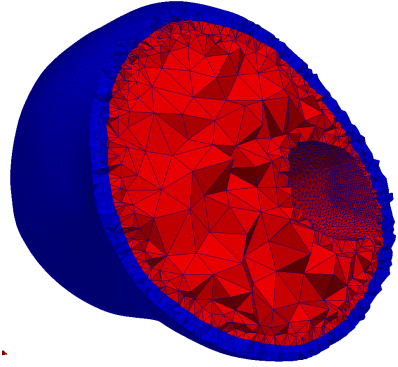

**Figure 2:** Example of finite element mesh used in this study. The AAA wall is meshed using two layers of elements (the number of layers is a configurable). The ILT is meshed using a minimum of two layers of elements (the number of layers is a configurable); the element size is increased in the middle of the ILT layer to reduce the number of elements in the mesh. To alleviate volumetric locking we used 10-noded parabolic tetrahedra. A typical mesh contains 500,000 elements.

Patient-specific median arterial pressure applied to the ILT surface was used as the loading condition, with ILT assumed to be 20 times more compliant than the AAA wall. The finite element simulations were carried out using the procedure described in the article by Joldes et al. (2016) published in *Journal of the Mechanical Behavior of Biomedical Materials*, which allows the computation of stress in the AAA wall without exact knowledge of the material properties. This is of great practical significance as patient-specific material properties for the AAA wall and ILT are currently impossible to obtain *in vivo*. For a detailed discussion of the problem of obtaining solutions without knowing mechanical properties of tissues, please see also Miller and Lu (2013) and Wittek et al. (2009).

The results of finite element simulations (maximum principal stresses in the AAA wall) are extracted by BioPARR for visualisation and analysis.

### 2.5   Incorporation of residual stress

Blood vessels in humans exhibit characteristics of a pre-stressed vessel i.e. one that is stressed, even when unloaded by external forces (Fung, 1991). This phenomenon is thought to be caused by biological remodelling of protein fibres and smoothing of muscle tone to reach a uniform stress state across the arterial wall thickness. This natural tendency of arteries to remodel towards a state which reduces stress concentrations is referred to in the literature as the Uniform Stress Hypothesis (Fung, 1991; Polzer et al., 2013).

These residual stresses are typically disregarded when analysing the AAA biomechanics. However, residual stresses have been shown to have a significant impact on the distribution of the wall stress (Raghavan and Vorp, 2000). In this study we account for residual stresses present in the aortic wall *in vivo*. Recently, Joldes et al. (2018) presented a new method for adding these residual stresses as a post-processing step. The method uses the Uniform Stress Hypothesis and seeks to average the wall stress over the wall thickness. This method gives consistent results that are comparable to existing iterative methods (Polzer et al., 2013). Crucially, the simplicity of this method leads to faster computations than existing alternatives and makes it compatible with existing clinical workflows (Joldes et al., 2018).

The entire analysis workflow is presented in Figure 3.

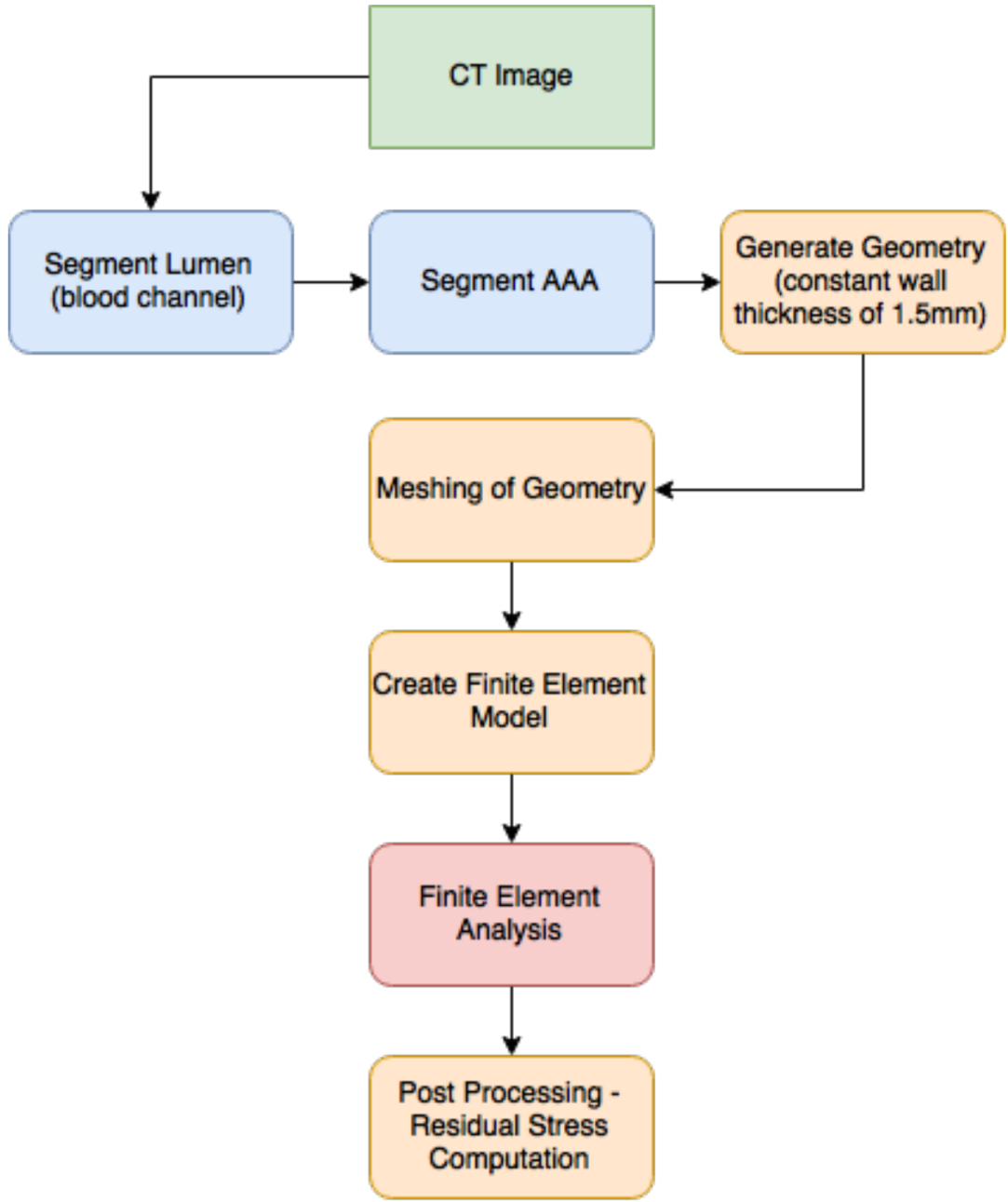

**Figure 3:** AAA analysis workflow using BioPARR. Steps in blue are performed in 3D Slicer (www.slicer.org). Step in red is performed in Abaqus finite element (FEM) code (other linear FEM solvers could be used). BioPARR performs the remaining steps (in orange) semi-automatically.

# 3    Results

We chose to use a 99th percentile maximum principal stress as a scalar indicator of the internal forces being withstood by the wall tissue.

The results for the 19 cases we analysed are given in Figures 4 and 5.

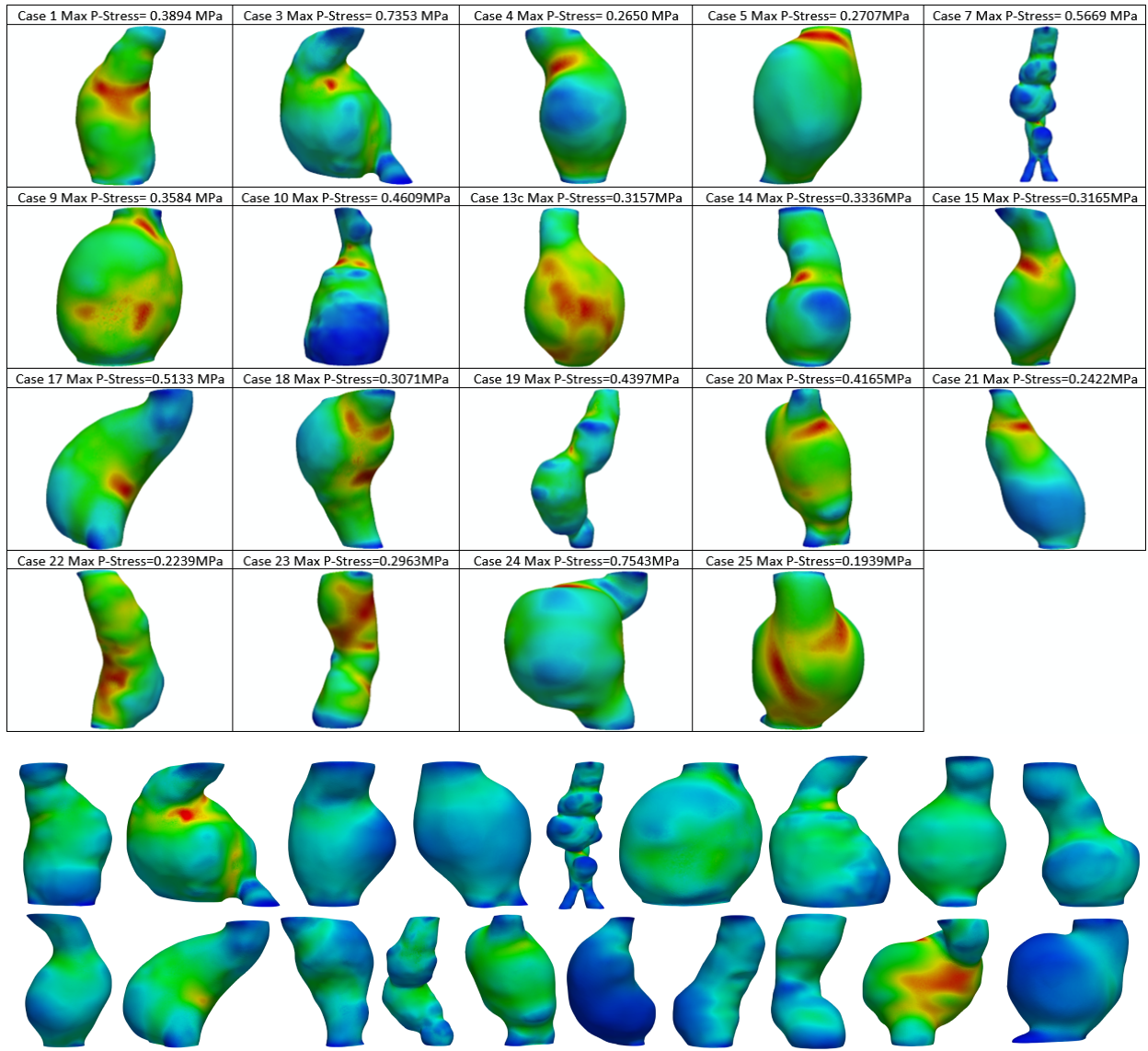

**Figure 4:** Contour plots of the stress distributions for 19 cases. As can be seen, no pattern emerges, and identifying symptomatic cases is not possible. Cases 21, 22, 23, 24 and 25 are symptomatic.

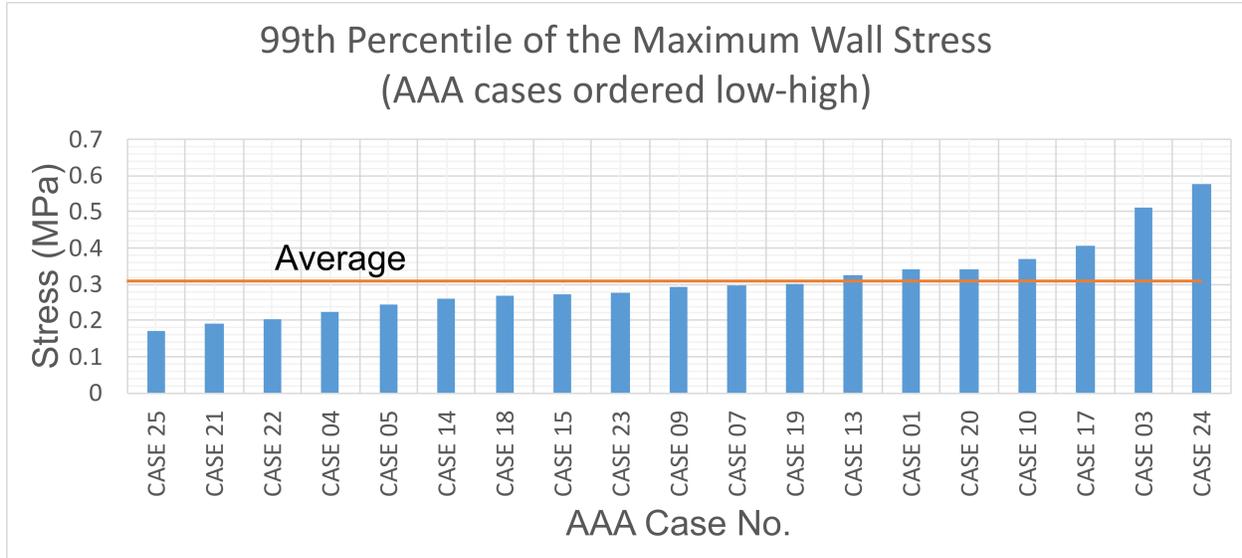

**Figure 5:** The AAA cases ordered from the lowest to the highest maximum wall principal stress (99[th] stress percentile). The stress is in MPa. No two separate (i.e. symptomatic vs asymptomatic) groups can be distinguished. Cases 21, 22, 23, 24 and 25 were symptomatic. Case 25 has the lowest and case 24 the highest principal stress.

## 4    Discussion and Conclusions

The results presented above suggest, somewhat surprisingly, that there is no correlation between the wall stress distribution or maximum principal stress values, and clinical observation of symptoms in patients with AAA. In the previous work, reviewed extensively vs Khosla et al. (2014), the patients being considered are those who have been operated on, which would include both symptomatic and asymptomatic cases. It is possible that this factor has contributed to reaching the conclusion that high stress in aortic wall correlates with the symptoms.

When considering the results we present here, one needs to ponder limitations of our modelling and simulation method. Firstly, due to limitations of resolution and contrast of clinical CT images used in this study we were unable to include patient-specific thickness of the AAA wall. Unfortunately most other studies suffer from the same deficiency. Nevertheless our recent results (Miller et al., 2019) suggest that maximum principal wall stress is proportional to the average wall thickness allowing a certain degree of optimism with regard to obtaining stress envelopes for a particular patient without the exact information about patient-specific wall thickness. Further limitation more relevant to the clinical use of these results is that we do not have a definitive value for the ultimate failure stress of the aortic wall. Nevertheless, compared with other methods of analysis available in the literature, our method embedded in BioPARR offers unparalleled simplicity and therefore compatibility with clinical workflows.

## Acknowledgements

The financial support of the National Health and Medical Research Council (Grant No. APP1063986) is gratefully acknowledged. We acknowledge the Raine Medical Research Foundation for funding G. R. Joldes through a Raine Priming Grant and support from Western Australian Department of Health.

We thank Ms Margot Chaix and Ms Laura Revol, visiting research students (from SIGMA-Clermont Graduate Engineering School in Clermont-Ferrand, France) at Intelligent Systems for Medicine Laboratory at the University of Western Australia, for help in formatting Figure 5.